\documentclass[12pt]{article}
\usepackage{amsmath}
\usepackage{amscd}
\usepackage{amssymb}
\usepackage{array}

\begin{document}
\rightline{CERN-TH/2002-314}
\rightline{UCLA/02/TEP/31}

\newcommand{\R}{\mathbb{R}}
\newcommand{\C}{\mathbb{C}}
\newcommand{\Z}{\mathbb{Z}}
\newcommand{\Hb}{\mathbb{H}}

\newcommand{\rE}{\mathrm{E}}
\newcommand{\rSp}{\mathrm{Sp}}
\newcommand{\rSO}{\mathrm{SO}}
\newcommand{\rSL}{\mathrm{SL}}
\newcommand{\rSU}{\mathrm{SU}}
\newcommand{\rUSp}{\mathrm{USp}}
\newcommand{\rU}{\mathrm{U}}
\newcommand{\rF}{\mathrm{F}}
\newcommand{\rGL}{\mathrm{GL}}
\newcommand{\rG}{\mathrm{G}}
\newcommand{\rK}{\mathrm{K}}

\newcommand{\fgl}{\mathfrak{gl}}
\newcommand{\fu}{\mathfrak{u}}
\newcommand{\fsl}{\mathfrak{sl}}
\newcommand{\fsp}{\mathfrak{sp}}
\newcommand{\fusp}{\mathfrak{usp}}
\newcommand{\fsu}{\mathfrak{su}}
\newcommand{\fp}{\mathfrak{p}}
\newcommand{\fso}{\mathfrak{so}}
\newcommand{\fl}{\mathfrak{l}}
\newcommand{\fg}{\mathfrak{g}}
\newcommand{\fr}{\mathfrak{r}}
\newcommand{\fe}{\mathfrak{e}}
\newcommand{\ft}{\mathfrak{t}}

\newcommand{\id}{\relax{\rm 1\kern-.35em 1}}
\vskip 1.5cm

  \centerline{\LARGE \bf Duality, gauging and superHiggs effect}
  
  \bigskip
  
  \centerline{\LARGE\bf in string and M-theory }

 \vskip 3cm
 \centerline{\large Sergio Ferrara}

\vskip 1.5cm

\centerline{\it CERN, Theory Division, CH 1211 Geneva 23,
Switzerland, }

\medskip

\centerline{\it Department of Physics and Astronomy,} \centerline{\it University of California,
Los Angeles. Los Angeles, USA, } 

\smallskip
\centerline{\it and} 

\smallskip

\centerline{\it  INFN, Laboratori Nazionali di
Frascati, Italy.}

\vskip 1cm

\begin{abstract}
We consider no-scale extended supergravity models as they arise from string and M-theory compactifications
 in presence of fluxes. The special role of gauging axion symmetries for the Higgs and superHiggs mechanism is 
outlined.
\end{abstract}

\vskip 2cm

\noindent Based on plenary talks presented at the String theory conferences in Hongzhou and Beijing, August 
2002,  China.
\vfill\eject

\section{Introduction}

Recently, supergravity models which admit a superHiggs mechanism  whithout tree 
level cosmological constant have been reconsidered in the framework of superstring 
and M-theory compactifications in presence of fluxes \cite{ps,drs,tv,ma,cklt,gkp,fp,kst}.

These models are a variant of the well known $AdS_5\times S^5$ compactification \cite{mal,gkpo,wi} of type IIB 
string theory, which in the 
supergravity limit \cite{krvn} corresponds to an SU(4) gauge $D=5$ maximally extended supergravity 
\cite{grw,ppvn}. The non abelian nature of the 
gauging is closely connected to the isometries of the 5-sphere and the gauge coupling is turned on by the 
5-form 
Ramond-Ramond flux. These theory has 32 unbroken supersymmetries in $D=5$.

Theories with less supersymmetries naturally arise in scenarios where turning on fluxes gives rise to 
compactifications 
without cosmological constant. 

These theories cover, in particular, the Scherk-Schwarz mechanism \cite{ss,css} as originally proposed in 
eleven dimensional 
supergravity and type IIB orientifolds in presence of three-form fluxes \cite{gkp,fp,kst}. In this class of 
theories fluxes correspond in 
the effective supergravity to charge couplings of axions to gauge fields
$$D_\mu \Phi^\Lambda=\partial_\mu\Phi^\Lambda + g_I^\Lambda A_\mu^I,$$
and the nature of the charge $g^\Lambda_I$ very much depends of the theory under consideration. 
The gauge transformations are

\begin{eqnarray*}
A_\mu^I\rightarrow A_\mu^I-\partial_\mu\xi^I,\\
\Phi^\Lambda\rightarrow \Phi^\Lambda +g_I^\Lambda \xi^I.\end{eqnarray*}

If the gauge group is non abelian then a possible situation is to have a ``flat group" and the latter is what 
occurs in 
the Scherk-Schwarz" mechanism. This is actually what is needed if the super Higgs effect involves BPS 
multiplets, which 
must be charged under central charges which are unbroken U(1) symmetries of the theory \cite{adfl1}.

Supersymmetry also requires that there exists a section $X$ on the scalar manifold such that 
$$g_I^\Lambda{X^I_\Lambda
}_{AB}$$
is a symmetric matrix in the  R-symmetry labels $A, B=1,\dots N$. The gravitino mass term has the form 
$$\bar \Psi_\mu^Ag_I^\Lambda{X^I_\Lambda}_{AB}\gamma^{\mu\nu}\Psi_\nu^\beta +\mathrm{h.c.}.$$

String inspired models with spontaneous supersymmetry breaking in flat space \cite{ckvpdfwg} have been analyzed 
in the past \cite{fkpz}. The superHiggs effect in the context of string theory \cite{bkl} and of brane scenarios 
for particle physics \cite{mp} have been explored more recently. 

In $N$-extended supergravity the scalar potential is generated by the gauge symmetry, and if charged scalar 
fields are 
present these symmetry must be an isometry of the scalar manifold. We will call such isometries U-dualities 
with an 
abuse of language which comes from string theory \cite{ht}.

Because of the particular nature of the theory, which couples the sigma models to vector fields, these 
isometries must 
have a particular action on the vectors and this is discussed in sections 2 and 3. In the subsequent sections, 
the 
examples of $N=8$ spontaneously broken supersymmetry \`a la Scherk-Schwarz and the $N=4$ theory coming from 
type IIB 
orientifolds will be discussed and their gauge theory structure emphasized. In the last section the role of 
translational 
isometries in the Higgs mechanism underlying the superHiggs effect will be considered in the special case when 
the 
scalar manifold is a symmetric space.

\section{Duality and supersymmetry breaking}

Extended supergravity theories enjoy beautiful covariance properties under duality rotations. Electric and 
magnetic field 
strengths undergo symplectic transformations
$$\begin{pmatrix} \mathcal{F}'\\\mathcal{G}'\end{pmatrix}=\begin{pmatrix} A&B\\C&D\end{pmatrix}
\begin{pmatrix} \mathcal{F}\\\mathcal{G}\end{pmatrix}$$
with $A^TD-C^TB=\id$, $A^TC$ and $B^TD$  symmetric.
The complexified coupling constant $\mathcal{N}_{\Lambda\Sigma}$, 
($\mathcal{G}^+_\Lambda=\mathcal{N}_{\Lambda\Sigma}\mathcal{F}^{+\Sigma}$) undergo fractional transformations 
\cite{cdfvp}
$$\mathcal{N}'=(C+D\mathcal{N})(A+B\mathcal{N})^{-1}.$$ In particular, if the theory has an invariance group G, 
it must 
have a symplectic action \cite{gz} on the pair $(\mathcal{F},\mathcal{G})$. We will generically call U-duality 
the invariance group of 
a given supergravity theory, that for our purposes of gauging must be a continuous group of symmetries. As a 
consequence of 
the gauging, the fermionic variations acquire the following terms \cite{df}
\begin{eqnarray*}\delta\Psi_{A\mu}&=&\cdots +\frac{S_{AB}}{2}\gamma_\mu\epsilon^B\\
\delta\lambda^I&=&\cdots +N^{IA}\epsilon_A\end{eqnarray*}
and the scalar potential is given by \cite{fm,cgp}
$$\delta_B^AV=-3\bar S^{AC}S_{CB} +N^{AI}N_{BI}, \qquad N_{BI}=(N^{BI})^*.$$
Then it follows ($A=B$) that
$$V=-3\sum_C\bar S^{AC}S_{AC}+N^{AI}N_{AI} \qquad \forall A$$
Flat space demands that 
$$3\sum_C\bar S^{AC}S_{AC}=\sum_IN^{AI}N_{AI} \qquad \forall A.$$
Furthermore, if the $A_0$-supersymmetry is unbroken, then 
$$\sum_C\bar S^{A_0C}S_{A_0C}=0=N^{A_0I}N_{A_0I}.$$

For spontaneously broken extended supergravity, a given residual supersymmetry requires a combination of the 
super Higgs 
and Higgs effects which give further constraints on the theory. 

For example, if a given theory breaks $N=2$ to $N=1$, there are two massive vector partners of the massive 
gravitino and 
then a Higgs effect must occur. It was shown in Ref. \cite{lo} that the condition for this to happen is that 
the scalar 
quaternionic manifold must have two translational isometries which get spontaneously broken. 

For the class of models under consideration, it is actually true that all massive vector particles, which 
undergo a Higgs 
mechanism as a consequence of supersymmetry breaking correspond to spontaneously broken translational isometries of 
the 
original non linear sigma model \cite{adfl2,adfl3}. 

For no-scale supergravity models \cite{cfkn,elnt}, the scalar potential is positive semidefinite. In that, the 
contribution to the 
potential comes only from those spin 1/2 fields which do not participate into the sueprsymmetry breaking.

\begin{eqnarray*}\langle \delta\chi^{\bar I}\rangle=N^{\bar I A}\epsilon_A=0\\
V=\sum_{\bar I} N^{\bar I A}N_{\bar I A}\end{eqnarray*}
while there is a cancellation of the $|S^{AB}|^2$ term with the spin 1/2 fermions for which 
$N^{IA}\epsilon_A\neq0$.

\section{Gauging of duality symmetries}

We consider  $D=4$ supergravity  in absence of fluxes. Let G be
the U-duality group of the theory and  $n$ the number of vectors
in the theory, $A_\mu^\Lambda,\; \Lambda=1\dots n$.  The field
strengths $F^\Lambda_{\mu\nu}$ and their duals
$G_\Lambda^{\mu\nu}=\partial\mathcal{L}/\partial
F^\Lambda_{\mu\nu}$ together carry a linear representation of  G.
When G is considered as embedded in $\rSp(2n, \R)$, the
representation carried by $\{F^\Lambda, G_\Lambda\}$ is promoted
to a representation of the full symplectic group \cite{gz}.

An arbitrary matrix of the  Lie algebra $\fsp(2n, \R)$ can be
written in terms of blocks of size $n\times n$
\begin{equation}\label{matrix}X=\begin{pmatrix}a&b\\c&-a^T\end{pmatrix},\qquad
b=b^T, \; c=c^T, \end{equation}
 and $a$ an arbitrary matrix of
$\fgl(n,\R)$. The Lie algebra $\fsp(2n, \R)$ admits then the
decomposition $$\fsp(2n,
\R)=\tilde\fg^0+\tilde\fg^{+1}+\tilde\fg^{-1},$$ with
  $$\tilde\fg^0=\bigr\{\begin{pmatrix}a&0\\0&-a^T\end{pmatrix}\bigl\}, \quad
\tilde\fg^{+1}=\bigr\{\begin{pmatrix}0&0\\c&0\end{pmatrix}\bigl\}
,\quad
\tilde\fg^{-1}=\bigr\{\begin{pmatrix}0&b\\0&0\end{pmatrix}\bigl\}.$$
So $\tilde\fg^0\approx\fgl(n,\R)$, $\tilde\fg^{+1}$ carries the
representation of $\fgl(n,\R)$
$\mathrm{sym}(\mathbf{n'}\otimes\mathbf{n'})$
 and $\tilde\fg^{-1}$  the representation
$\mathrm{sym}(\mathbf{n}\otimes\mathbf{n})$. $\mathbf{n'}$ denotes
the contragradient representation of $\mathbf{n}$. The subalgebra
$\fso(1,1)_z\subset \fgl(n,\R)$ (the subindex $z$ is to
distinguish it from other subalgebras $\fso(1,1)$ that we will
consider in the following), whose generator is the element
$-\frac{1}{2}\id$, acts with charge $+1$ on $\tilde\fg^{+1}$ and
with charge $-1$ on $\tilde\fg^{-1}$ (so the upper indices
indicate this charge). In fact, $\fso(1,1)_z$ defines a grading of
$\fsp(2n, \R)$ and consequently $\tilde\fg^{\pm 1}$ are abelian
subalgebras.
 The matrices of the subalgebra $\tilde\fg^0+\tilde\fg^{+1}$ have lower
block-triangular form ($b=0$). The vector space carrying the
fundamental representation of the symplectic algebra inherits also
a grading and it decomposes as $V=V^+ \oplus V^-$. Notice that
$V^+$ carries a representation of $\tilde\fg^0+\tilde\fg^{+1}$.

We consider now the U-duality group G with Lie algebra $\fg\subset
\fsp(2n)$. Any subalgebra of $\fg$ which is a subalgebra of the
lower triangular matrices $\tilde\fg^0+\tilde\fg^{+1}$ will
transform the field strengths (in $V^+$)  without involving their
magnetic duals. Then, the vector potentials themselves carry a
linear representation of this subalgebra which will be called an
{\it electric subalgebra} of $\fg$ (and generically denoted by
$\fg_{\mathrm{el}}$). The corresponding group,
$\rG_{\mathrm{el}}$, is an {\it electric subgroup} of G. The gauge
group  $\rG_{\mathrm{gauge}}$ is a subgroup of the
$\rG_{\mathrm{el}}$ such that its action on the vector potentials
is the adjoint action. As we will see, there is not a unique
maximal electric subgroup of G, and this gives rise to many
different gaugings of the supergravity theory.
		
In the next sections, we will discuss the examples of $N=8,4,$
supergravity from this new point of view.

\section{$N=8$ supergravity\label{N=8}}

The U-duality group of $N=8$ supergravity is $\rG=\rE_{7,7}$
\cite{cj}, which can be embedded in several ways in $\rSp(56,\R)$
(there are 28 vector fields). The electric subalgebras will always
be subalgebras of $\fg_{\mathrm{el}}\subset
\tilde\fg^0+\tilde\fg^{+1}$, being in this case
$\tilde\fg^0\approx \fsl(28, \R ) + \fso(1,1)_z$.

\bigskip
 Consider the following
decomposition of $\fe_{7,7}$ \begin{equation}\fe_{7,7}=\fsl(8,\R)
+ \mathbf{70} \label{e77}.\end{equation} $\fsl(8,\R)$ is a maximal
subalgebra of $\fe_{7,7}$ and $ \mathbf{70}$ is an irreducible
representation of $\fsl(8,\R)$, the four-fold antisymmetric.

The representation ${\bf 56}$ of $\fe_{7,7}$ decomposes under the
subgroup $\fsl(8,\R)$ as $$\mathbf{56}\longrightarrow
\mathbf{28}+\mathbf{28'},$$ where $\mathbf{28}$ and $\mathbf{28'}$
are two-fold antisymmetric representations of $\fsl(8,\R)$.

The embedding $\fe_{7,7}\subset \fsp(56,\R)$ is constructed as
follows. We have that $\fsl(8,\R)\subset \fsl(28,\R)$ by means of
the two-fold antisymmetric representation (the $\mathbf{28}$) of
$\fsl(8,\R)$.
 With
a  two-fold antisymmetric tensor we can construct a four-fold
antisymmetric tensor by taking the symmetrized tensor product. In
this way the generators in the
 {\bf 70} of (\ref{e77}) are realized in the two-fold symmetric
 representation of $\fsl(28,\R)$, $\mathrm{sym}(\mathbf{28}\otimes \mathbf{28})$,
 forming the $b$ matrix of (\ref{matrix}),
 $b_{\{[AB][CD]\}}$. Since in $\fsl(8,\R)$ there is
 an invariant, totally antisymmetric tensor $\epsilon_{A_1\dots
 A_8}$, we have another symmetric matrix
 $$c^{\{[A_1A_2][A_3
 A_4]\}}=\frac 1{4!}\epsilon^{A_1\dots
 A_8}b_{\{[A_5A_6][A_7
 A_8]\}}.$$

 This is the standard
embedding of $\fe_{7,7}$ in $\fsp(56,\R)$.  $\fsl(8,\R)$ is a
maximal electric subalgebra. The gauging of different electric
subalgebras of $\fsl(8,\R)$ and of its contractions gives rise to
all the theories described in \cite{hw,cfgtt}. In this choice,
SO(8) is the maximal compact electric subgroup.

\bigskip

 We will now consider a different embedding, which is
the one relevant for the Scherk--Schwarz mechanism \cite{adfl2}.  Consider the
decomposition $$\fe_{7,7}=\fe_{6,6}+\fso(1,1)_{k} +
\mathbf{27_{-2}}+\mathbf{27'_{+2}},$$ where
$\fe_{6,6}+\fso(1,1)_{k}$ is a maximal reductive
subalgebra. Notice that it is not a
maximal subalgebra.  In five dimensions the U-duality group is
$\rE_{6,6}$ and it is totally electric. If we want to see the four
dimensional theory as the dimensional reduction of a five
dimensional one, this is the natural decomposition to consider,
and the $\fso(1,1)_{k}$ rescales the modulus of the
compactification radius of the 5th dimension. The normalization of
the generator of $\fso(1,1)_{k}$ has been chosen in such way that
the fundamental representation decomposes as
$$\mathbf{56}\rightarrow \mathbf{27_{+1}}
+\mathbf{1_{+3}}+\mathbf{27'_{-1}}+\mathbf{1_{-3}}. $$ (Note that
the ratio one to three of the $\fso(1,1)_{k}$ charges is what one
obtains for the relative charges of the 27 five-dimensional
vectors versus the graviphoton in the standard Kaluza--Klein
reduction).

 We have that $\fe_{6,6}+\fso(1,1)_{k}
\subset\fgl(28,\R)$, by decomposing the fundamental representation
$$\mathbf{28}\rightarrow \mathbf{27_{+1}}+\mathbf{1_{+3}}$$ but
$\fso(1,1)_{\mathrm{k}}$ does not correspond to the trace
generator $\fso(1,1)_z$ in $\fgl(28,\R)$, since all the  vectors
in the representation $\mathbf{28}$  have  the same charge under
$\fso(1,1)_z$, $\mathbf{27_z}+\mathbf{1_z}$. Indeed there is
another subalgebra $\fso(1,1)_r$ in $\fgl(28,\R)$ which commutes
with $\fe_{6,6}$. This comes from the sequence of embeddings
$$\fe_{6,6}\subset\fsl(27,\R)\subset
\fgl(27,\R)=\fsl(27,\R)+\fso(1,1)_r\subset \fsl(28,\R)$$
corresponding to the fact that only 27 of the 28 vectors are
transformed by $\fe_{6,6}$. The charges of the 28 vectors are
$\mathbf{27_r}+\mathbf{1_{-27r}}$. Then $\fso(1,1)_{k}$ turns out
to be a combination of $\fso(1,1)_z$ and $\fso(1,1)_r$.

In this setting the symplectic embedding of $\fe_{7,7}$ in
$\fsp(56,\R)$ is different from the standard one previously
considered. It was explicitly worked out in \cite{adfl2}. Here
$\fg^0 = \fe_{6,6}+\fso(1,1)_{k}$ is the block diagonal part and
$$\fg_{\mathrm{el}} = \fe_{6,6}+\fso(1,1)_{k} +\mathbf{27'_{+2}}$$
is lower block-triangular. Then, it is an electric subalgebra.
\par
Note that in order to have a physical theory the unbroken gauge
group in the $\fg^0$ part must belong to the maximal compact
subgroup of $\fg^0$.

The Scherk--Schwarz \cite{ss,css} mechanism corresponds to the
gauging of an electric subgroup (a ``flat group") with algebra
$\fg_{\mathrm{el}}=\fu(1)\circledS\mathbf{27'_{+2}}$ (semidirect
sum), where
 $\fu(1)$ is a generic element of the Cartan
subalgebra of the maximal compact subgroup $\fusp(8)$ of
$\fe_{6,6}$ \cite{adfl2}. The gauging of this electric group
breaks spontaneously the supersymmetry. Partial breaking is
allowed, and the unbroken supersymmetry algebra has a central
charge.  Central charges are $\fu(1)$ symmetries which belong to
the CSA of G, so if they belong to $\fg_{\mathrm{el}}$ they must
belong to  the maximal compact subalgebra of $\fg^0\subset
\fg_{\mathrm{el}}$. In our case we have one central charge which
is identified with the $\fu(1)$ factor in the semidirect sum
$\fg_{\mathrm{el}}$.

In fact, the Scherk--Schwarz mechanism allows partial breakings
$N=8\rightarrow N'=6,4,2,0$, and the spin 3/2 multiplets are 1/2
BPS, which means that only one central charge is present. The
number of unbroken translational symmetries in the phases
$N'=6,4,2,0$ is, respectively, 15,7,3,3.

\section{$N=4$ supergravity. \label{N=4}}

 As we are going to see, a richer structure emerges in the $N=4$ theory
 because in this case
we can have both non abelian and abelian flat groups, depending on
the particular model we consider.

Let us consider the $N=4$ theory with $n_v+1$ vector multiplets.
The U-duality group is $\rG=\rSO(6,n_v+1)\times\rSL(2,\R)$,
embedded  (in different ways) in  $\rSp(2(6+n_v+1),\R)$, so any
electric subalgebra must have block diagonal part a subalgebra
$\fg^0 \subset \fsl(6+n_v+1,\R) \times \fso(1,1)_z$.

\bigskip

The standard embedding corresponds to take
$$\fso(6,n_v+1)+\fso(1,1)_q\subset \fgl(6+n_v+1,\R).$$
 Here
$\fso(1,1)_q$ is the Cartan subalgebra of $\fsl(2,\R)$ and it is
identified with $\fso(1,1)_z$. The only off-diagonal elements are
the other two generators of $\fsl(2,\R)$, say $X^\pm$. The
electric subalgebra is then the lower triangular subalgebra
$(\fso(6,n_v+1)+\fso(1,1)_q)\circledS \{X^+\}$. This embedding
appears when doing the  compactification of the heterotic string
on $T^6$. (See for example the review of Ref. \cite{gpr}).

\bigskip

We analyze now another symplectic embedding. We take  $n_v=5$ (the
embedding is possible only for $n_v\geq 5$). Then we have the
following decomposition
 $$\fso(6,6)=
\fsl(6,\R)+ \fso(1,1)_s+\mathbf{{15'}^+}+\mathbf{{15}^-}, $$ where
$\mathbf{15}$ is the two-fold antisymmetric representation. Since
$\fsl(n)+\fsl(m)\subset \fsl(nm)$, we have that
$$\fsl(6,\R)+\fsl(2,\R)+\fso(1,1)_s\subset \fgl(12,\R).$$ The
representation $(\mathbf{15'},\mathbf{1})$ is symmetric (the
singlet of $\fsl(2,\R)$ is the two-fold antisymmetric), so we have
that $\mathbf{{15'}^+}\subset\tilde\fg^+\subset \fsp(24,\R)$. This
defines the symplectic embedding. The $\fso(1,1)_s$ is identified
with $\fso(1,1)_z$ of the symplectic algebra.

 The representation $\mathbf{12}$ of
$\fso(6,6)$ decomposes, with respect to $\fsl(6,\R)+\fso(1,1)_z$,
as $$\mathbf{12}\rightarrow \mathbf{6_{+1}}+\mathbf{6_{-1}},$$
thus containing six electric and six magnetic fields, and the
bifundamental of $\fso(6,6)+\fsl(2,\R)$ decomposes as
$$\mathbf{(12,2)}=\mathbf{(6_{+1},2)}_{\mathrm{electric}}+\mathbf{(6_{-1},2)}_{\mathrm{magnetic}}.$$
In particular, we see that $\fsl(2,\R)$ is totally electric.

The twelve vectors gauge an abelian 12-dimensional subgroup of the
$\mathbf{{15'}^+}$ translations.

This model was investigated in  Ref. \cite{tz}, but from our point
of view  it comes from the general analysis of gauging flat
groups. In this case the flat group is completely abelian, since
no central charge is gauged. The theory has four independent mass
parameters, the four masses of the gravitinos. This allows  a
partial supersymmetry breaking without cosmological constant from
$N=4\rightarrow N'=3,2,1,0$, where the massive gravitinos belong
to long (non BPS) massive representations in all the cases, as it
is implied by the fact that the central charge is not gauged and
the fields are not charged under it.

  From the analysis of the
consistent truncation of $N=4\rightarrow N'=3$ supergravity, it is
known that $N=4$ supergravity coupled to 6 matter vector
multiplets can indeed be consistently reduced to an $N=3$ theory
coupled to 3 matter multiplets \cite{adfl1}. Correspondingly we
have, for the scalar manifolds of such models,
$$\rSU(3,3)/(\rSU(3)\times\rSU(3)\times\rU(1))\subset\rSO(6,6)/(\rSO(6)\times\rSO(6)).$$
In fact,   the Higgs effect in this theory needs the gauging of a
group of dimension 12, spontaneously broken to a group of
dimension 6.
 The
scalar manifolds of the broken phases are
\begin{eqnarray*}& \rSU(3,3)/(\rSU(3)\times\rSU(3)\times\rU(1)),\quad
&\mathrm{for}\quad N'=3\\&
(\rSU(1,1)/\rU(1))\times\rSU(2,2)/(\rSU(2)\times\rSU(2)\times\rU(1)),\quad
&\mathrm{for}\quad N'=2\\& (\rSU(1,1)/\rU(1))^3,\quad
&\mathrm{for}\quad N'=1.\end{eqnarray*}  These models are also
analyzed in Ref. \cite{fp,kst} from another point of view. There,
Type IIB supergravity is compactified on the $T_6/\Z_2$
orientifold with brane fluxes turned on and the same pattern of
spontaneous symmetry breaking is found.

\section{Translational symmetries, Goldstone bosons and Higgs effect}

When the manifold parametrized by the scalars is a coset space
$G/H$, there is an abelian  algebra of isometries that is
contained in $G$. This algebra is the maximal abelian ideal of the
solvable algebra associated to the coset space via the Iwasawa
decomposition (see for example \cite{he}).  In the examples that
we analyze in this section we have two models with two coset
spaces, $G/H$ corresponding to the unbroken supersymmetry model
and $G'/H'$ corresponding to the model with partial breaking of
supersymmetry once  the massive modes have been integrated out.
We will denote by $\ft(G/H)$ and $\ft'(G'/H')$ the abelian
subalgebras associated to the respective cosets (here the
``$\ft$'' stands for translational). $t$ and $t'$ are respectively
the dimensions of these  subalgebras.

If $n_v$ and $n_v'$ denote the number of massless  vectors in each
theory, we find  in  all the models analyzed that $t-t'=n_v-n_v'$.
The solvable group obtained in the Iwasawa decomposition (now in
the group instead that in the algebra) is diffeomorphic as a
manifold to $G/H$.  This parametrization has been considered in
the literature to analyze U-dualities in string theory
\cite{adfft,lps,cj}. The generators of the maximal abelian ideal
act as  translations on $t$ of the coordinates of $G/H$, which
appear only through derivatives in the Lagrangian and
 which are flat directions of  the scalar potential. This suggests that, as a general
rule for a consistent Higgs effect, these particular coordinates
are the Goldstone bosons connected to the spontaneous breaking of
$\R^{n_v}$ to $\R^{n_v'}$, so they have been absorbed into the
vectors that have acquired mass.

We first analyze first  that are obtained with the
Scherk-Schwarz mechanism. In  these cases one can prove that the
above considerations are actually valid \cite{adfl1}. It
would be interesting to know the cases where this rule does not
hold.

\subparagraph{$N=8\rightarrow N'=6$.} The coset space of the
scalars in  $N=8$ supergravity is $\rE_{7,7}/\rSU(8)$ and the
dimension of the translational subalgebra is $t=27$ \cite{adfft}.
In $N'=6$ the coset is $\rSO^*(12)/\rU(6)$, and $t'=15$. So
$t-t'=12$. It is easy to see that $n_v-n_v'=28-16=12$.

\subparagraph{$N=8\rightarrow N'=2$.} For $N'=2$ we  have a
certain number of vector multiplets  ($n_1$) and hypermultiplets
($n_2$).

The minimal model($m_i\neq m_j,\; i,j= 2,3,4$ in the notation of
Ref. \cite{ss,css} corresponds to  $n_1=3$ (we take $n_2=0$),
and  the coset is
$$\frac{\rSU(1,1)}{\rU(1)}\times\frac{\rSU(1,1)}{\rU(1)}\times\frac{\rSU(1,1)}{\rU(1)}.
$$$t'=3$, so $t-t'=27-3=24$. We have that  $n_v-n_v'=28-4=24$.

The maximal model ($m_1= m_2=m_3$) corresponds to having $n_1=9$, (again we take $n_2=0$). The coset space is
$$\frac{\rSU(3,3)}{\rSU(3)\times\rSU(3)\times\rU(1)}.$$ In this case $t-t'=27-9=18$ and $n_v-n_v'=28-10=18$.

\bigskip

The examples that follow can be obtained  in Type IIB superstring
 compactified on an orientifold $T^6/\Z_2$ in presence of brane fluxes  \cite{fp,kst} and in certain
  gauged supergravity theories \cite{tz,dfv}.
We  want to consider the spontaneous breaking of $N=4,3$
supergravities down to $N'=3,2$. In order to have a consistent
reduction it is necessary that  the scalar manifold of
 the broken theory is a submanifold of the unbroken one \cite{adf}. For the $N'=2$ case this is just
 an assumption since the effects of integrating out the massive modes could be more complicated.

\subparagraph{$N=4\rightarrow N'=3$.} We consider the $N=4$ model with six massless vector multiplets,

$$\frac{\rSO(6,6)}{\rSO(6)\times\rSO(6)}\times
\frac{\rSU(1,1)}{\rU(1)}, $$ with $t=15$. Note that the SU(1,1)
factor is not considered because SU(3,3)$\subset$SO(6,6) The
decomposition of the massless multiplets is

\begin{eqnarray*}\bigl[(2), 4(\frac{3}{2}), 6(1), 4(\frac{1}{2}),2(0)\bigr]+
6\bigl[(1), 4(\frac{1}{2}), 6(0)\bigr]\rightarrow\\ \bigl[(2),
3(\frac{3}{2}), 3(1), 1(\frac{1}{2})\bigr]+6\bigl[(1),
4(\frac{1}{2}),6(0)\bigr]\end{eqnarray*} In $N=3$ the long spin
3/2 multiplet is formed by adding 3 massless vector  multiplets to
the $\lambda_{MAX}=3/2$ multiplets. There remain three massless
vector multiplets. The scalar manifold of the theory is

$$\frac{\rSU(3,3)}{\rSU(3)\times\rSU(3)\times\rU(1)} $$
with $t'=9$. So we have $t-t'=15-9=6$ an $n_v-n_v'=12-6=6$.

\subparagraph{$N=3\rightarrow N'=2$.} We start with the $N=3$
model with 3 vector multiplets as above. The decomposition of the
graviton multiplet is

\begin{eqnarray*}\bigl[(2), 3(\frac{3}{2}), 3(1), (\frac{1}{2})\bigr]\rightarrow
\bigl[(2), 2(\frac{3}{2}), (1)\bigr]+\bigl[(\frac{3}{2}), 2(1),
(\frac{1}{2})\bigr]\end{eqnarray*} and the decomposition of the
massless vector multiplet is
\begin{eqnarray*}\bigl[(1), 4(\frac{1}{2}, 6(0))\bigr]\rightarrow
\bigl[(1), 2(\frac{1}{2}), 2(0)\bigr]+\bigl[2(\frac{1}{2}),
4(0)\bigr].\end{eqnarray*} To form a long spin 3/2 multiplet we
need two massless vector multiplets and one hypermultiplet.  The
residual theory has then one vector multiplet and two
hypermultiplets. The coset is then
$$\frac{\rSU(1,1)}{\rU(1)}\times\frac{\rSU(2,2)}{\rSU(2)\times\rSU(2)\times
\rU(1) } $$ with $t'=5$. So we have $t-t'=9-5=4$ and
$n_v-n_v'=6-2=4$.

\section*{Acknowledgments}
This report is based on  collaborations with L. Andrianopoli, R. D'Auria, M. A. Lled\'o and S. Vaul\'a, who 
are kindly acknowledged.

 Work supported in part by the European Community's Human
Potential Program under contract HPRN-CT-2000-00131 Quantum
Space-Time, and D.O.E. grant DE-FG03-91ER40662, Task C.


\begin{thebibliography}{99}

\bibitem{ps}
J. Polchinski and A. Strominger. {\it Phys.  Lett.  B} {\bf 388}, 736 (1996).
\bibitem{drs}
K. Dasgupta, G. Rajesh and S. Sethi. {\it JHEP} {\bf 9908}, 023 (1999).

\bibitem{tv}
T. R. Taylor and C. Vafa. {\it Phys. Lett. B} {\bf 474}, 130
(2000).


\bibitem{ma}
P. Mayr. {\it Nucl.  Phys.  B} {\bf 593}, 99
(2001).

\bibitem{cklt}
G. Curio, A. Klemm, D. Lust and S. Theisen. {\it Nucl. Phys. B} {\bf 609}, 3 (2001).


\bibitem{gkp} S. B. Giddings, S. Kachru and J. Polchinski.
hep-th/0105097.

\bibitem{fp}
A. R. Frey and J. Polchinski. {\it Phys. Rev.}  {\bf 65} 126009 (2002).


\bibitem{kst}
S. Kachru, M. Schulz and S. Trivedi.
 hep-th/0201028.
 
 \bibitem{mal} J. Maldacena, {\it Adv. Theor. Math. Phys.} {\bf 2}, 231 (1998).
 
 \bibitem{gkpo} S. Gubser, I. Klebanov and A. Polyakov, {\it  Phys. Lett. } {\bf B428}, 105 (1998).
 
 \bibitem{wi} E. Witten, {\it Adv. Theor. Math. Phys.} {\bf 2}, 253 (1998).
 
 
 
 
 
 \bibitem{krvn} H. J. Kim, L. Romans and P. van Nieuwenhuizen, {\it  Phys. Rev.} {\bf B32}, 389 (1985).
 
 \bibitem{grw} M. Gunaydin, L. Romans and N. P. Warner, {\it  Phys. Lett. } {\bf 154B}, 268 (1985).
 
 \bibitem{ppvn}l M. Pernici, K. Pilch and P. van Nieuwenhuizen, {\it  Nucl. Phys. } {\bf B259}, 460 (1985).
 
 \bibitem{ss}
J. Scherk and J. H. Schwarz, ``How To Get Masses From Extra
Dimensions''. {\it Nucl. Phys. B} {\bf 153}, 61 (1979).
 
\bibitem{css} E. Cremmer, J. Scherk and J. H. Schwarz.
{\it Phys. Lett. B} {\bf 84}, 83 (1979).
 
 
\bibitem{adfl1} L. Andrianopoli, R. D'Auria, S. Ferrara and M. A. Lled\'o. {\it  Nucl. Phys. } {\bf B640}, 46 
(2002).

\bibitem{fkpz} S. Ferrara, C. Kounnas, M. Porrati and F. Zwirner. {\it  Nucl. Phys. } {\bf B318}, 75 
(1989);{\it Phys. Lett. } {\bf B194} 366 (1987).



\bibitem{bkl}
R. Blumenhagen, C. Kounnas and D. Lust. {\it JHEP}
{\bf 0001}, 036 (2000);

I. Antoniadis, J. P. Derendinger and C. Kounnas. hep-th/9908137.


 E. Kiritsis and C. Kounnas.
{\it Nucl. Phys. B }{\bf 503} 117 (1997).





\bibitem{mp}
E.A. Mirabelli and M.E. Peskin. {\it Phys. Rev. D} {\bf
58},  065002 (1998);

R. Barbieri, L.J. Hall, Y. Nomura.  {\it
Nucl. Phys. B} {\bf 624}, 63  (2002);



 I. Antoniadis and M. Quir\'os. {\it Phys. Lett. B} {\bf 416}, 327 (1998);


J. Bagger, F. Feruglio and F. Zwirner. {\it JHEP} {\bf 0202} 010 (2002).



\bibitem{ckvpdfwg}
E. Cremmer, C. Kounnas, A. Van Proeyen, J. P. Derendinger, S.
Ferrara, B. de Wit and L. Girardello. {\it  Nucl. Phys. B} {\bf 250} 385 (1985).
 
 
 \bibitem{ht} C. M. Hull, P. K. Townsend, {\it  Nucl. Phys. } {\bf B451}, 525 (1995).
 
 \bibitem{cdfvp} A. Ceresole, R. D'Auria, S. Ferrara, A. van Proeyen. {\it  Nucl. Phys. } {\bf B444}, 92 (1995)
 
 \bibitem{gz}
M.K. Gaillard and B. Zumino. {\it Nucl. Phys. B} {\bf 193}, 221 (1981).

\bibitem{df}  R. D'Auria and S. Ferrara. {\it JHEP} {\bf 0105}, 034 (2001).


\bibitem{fm}
S. Ferrara and L. Maiani. Proceedings of SILARG V. Baricloche, Argentina. O. Bressan, M. Castagnino and V. 
Hamity editors. World Scientific (1985).

\bibitem{cgp} S. Cecotti, L. Girardello, M. Porrati. {\it  Nucl. Phys. } {\bf B268}, 295 (1986).

\bibitem{lo}
 J. Louis. hep-th/0203138.
 
 

\bibitem{adfl2} L. Andrianopoli, R. D'Auria, S. Ferrara and M. A. Lled\'o, {\it JHEP} {\bf 0207} 010 (2002).
hep-th/0204145.

\bibitem{adfl3} L. Andrianopoli, R. D'Auria, S. Ferrara and M. A. Lled\'o,
{\it  Nucl. Phys. } {\bf B640}, 63 (2002).



\bibitem{cfkn}
E. Cremmer, S. Ferrara, C. Kounnas and D. V. Nanopoulos.
{\it Phys. Lett. B} {\bf 133} 61 (1983).

\bibitem{elnt}
J. R. Ellis, A. B. Lahanas, D. V. Nanopoulos and K. Tamvakis. {\it  Phys. Lett. B }
{\bf 134} 429 (1984).

\bibitem{cj}
E. Cremmer and B. Julia. {\it Nucl.
Phys. B} {\bf 159}, 141 (1979).


\bibitem{hw}
C. M. Hull and N. P. Warner. {\it Nucl.  Phys. B} {\bf 253}, 650
(1985).


\bibitem{cfgtt}
F. Cordaro, P. Fre, L. Gualtieri, P. Termonia and M. Trigiante. {\it
Nucl. Phys. B } {\bf 532}, 245 (1998).

\bibitem{gpr}
A. Giveon, M. Porrati and E. Rabinovici. {\it  Phys. Rep.}  {\bf 244}, 77 (1994).



\bibitem{tz}
V. A. Tsokur and Y. M. Zinovev, {\it Phys.  Atom.
Nucl.}  {\bf 59}, 2192 (1996); {\it Phys.  Atom.  Nucl.}
{\bf 59} 2185 (1996).

\bibitem{dfv} R. D'Auria, S. Ferrara and  S. Vaul\'a, {\it New J.Phys} {\bf 4} 71 (2002);

R. D'Auria, S. Ferrara, M. A. Lled\'o  and  S. Vaul\'a,  hep-th/0211027.






 
\bibitem{he} S. Helgason, ``Differential Geometry, Lie Groups and Symmetric Spaces''. Academic Press, (1978).


\bibitem{adfft}
L. Andrianopoli, R. D'Auria, S. Ferrara, P. Fr\'e and M.
Trigiante.
{\it Nucl.  Phys.  B}  {\bf 496} 617 (1997);

L. Andrianopoli, R. D'Auria, S. Ferrara, P. Fr\'e, R. Minasian and
M. Trigiante. {\it Nucl. Phys. B} {\bf 493} 249 (1997).

\bibitem{lps}
H. Lu, C. N. Pope and K. S. Stelle. {\it Nucl.  Phys. B} {\bf 476} 89 (1996).

\bibitem{adf}
L. Andrianopoli, R. D'Auria and S. Ferrara.
{\it JHEP} {\bf 0203} 025(2002).















\end{thebibliography}
\end{document}